\documentclass{iau}
 \usepackage{graphicx} 

\title[IAUS291.~~Searches for continuous gravitational waves] 
{Searches for continuous gravitational waves with the LIGO and Virgo detector } 

\author[K. Riles]
{Keith Riles$^1$ \\for the LIGO Scientific Collaboration and Virgo Collaboration}  

\affiliation{$^1$Physics Department, Univ. Michigan, 450 Church St., Ann Arbor, MI 48109-1040, U.S.A.\\ email: {\tt kriles@umich.edu}
}

\pubyear{2012}
\volume{291}  
\jname{\mbox{Neutron Stars and Pulsars: Challenges and Opportunities after 80 years}}
\editors{J. van Leeuwen, ed.} 
\begin{document}

\maketitle

\begin{abstract}
The LIGO Scientific Collaboration and Virgo Collaboration have carried out joint searches in LIGO and Virgo data for periodic continuous gravitational waves. These analyses range from targeted searches for gravitational-wave signals from known pulsars, for which precise ephemerides from radio or X-ray observations are used in matched filters, to all-sky searches for unknown neutron stars, including stars in binary systems. Between these extremes lie directed searches for known stars of unknown spin frequency or for new unknown sources at specific locations, such as near the galactic center or in globular clusters. Recent and ongoing searches of each type will be summarized, along with prospects for future searches using data from the Advanced LIGO and Virgo detectors.
\end{abstract}


The hunt for gravitational waves has been a long one, carried out over
decades, using instruments ranging from the Earth itself (normal mode excitation),
to 1-ton metal bars to multi-km interferometers, to radio telescopes monitoring
the precise timing of arrays of millisecond pulsars scattered across the sky.
From the timing variations observed in the famous Hulse-Taylor binary pulsar
system (\cite{bib:hulsetaylor}), one can be confident that gravitational waves 
{\it are} emitted with appreciable magnitude by astrophysical systems, but
direct detection of those waves has proven to be an experimental challenge (see \cite{bib:rilesreview}).

The focus of this article is the search in data from the LIGO (see \cite{bib:ligo}) and 
Virgo (see \cite{bib:virgo})
laser interferometers for evidence of continuous gravitational waves in
the audio band, as might be radiated by nearby, rapidly spinning neutron stars.
To emit gravitational waves, a spinning star must be characterized by some
non-axisymmetry, \eg, due to a ``mountain'' ($\sim$mm high) at the star's equator
or due to stellar precession. Because resulting strain amplitudes reaching the Earth 
are expected to be
quite small ($\sim10^{-24}$ or much smaller), detection requires integration of
data streams over long observation spans (the longest to date being $\sim$23 months).

Because of computational cost considerations, it is natural to classify searches
for continuous gravitational waves into three broad categories: 1) targeted -- in which
precise pulsar ephemerides from radio, X-ray or $\gamma$-ray observations can be used in
a coherent integration over the full observation span; 2) directed -- in which
the direction of the source is known precisely, but for which little or no frequency
information is known; and 3) all-sky -- in which there is no information about location
or frequency. 

Targeted searches for known pulsars have been published based on data from the
first five LIGO data runs (S1--S5) and on the second Virgo data run (VSR2), with
continuing searches in the sixth LIGO run (S6) and the most recent Virgo
run (VSR4). 
The S5 search determined upper limits~(\cite[Abadie \etal\ 2010]{bib:cwtargeteds5}; based on 23 months of observation span) 
for 116 known pulsars in the LIGO band.
Highlights from the search
included a lowest-strain upper limit of $2.3\times10^{-26}$ (J1603-7202) and
a lowest stellar ellipticity limit of $7\times10^{-8}$ (J2124-3358). In addition,
a limit of 2\% was placed on the fraction of rotational energy loss of the 
Crab pulsar that can be attributed to gravitational radiation. A more recent
search (see \cite{bib:cwtargetedvela}) in Virgo VSR2 data for the Vela pulsar at an expected gravitational
wave frequency of about 22 Hz (for which Virgo sensitivity is substantially
better than LIGO's) yielded an upper limit of about 35\%\ on Vela's fractional
energy loss due to gravitational waves.

\begin{figure}[t!]
\begin{center}
\includegraphics[height=3.5in]{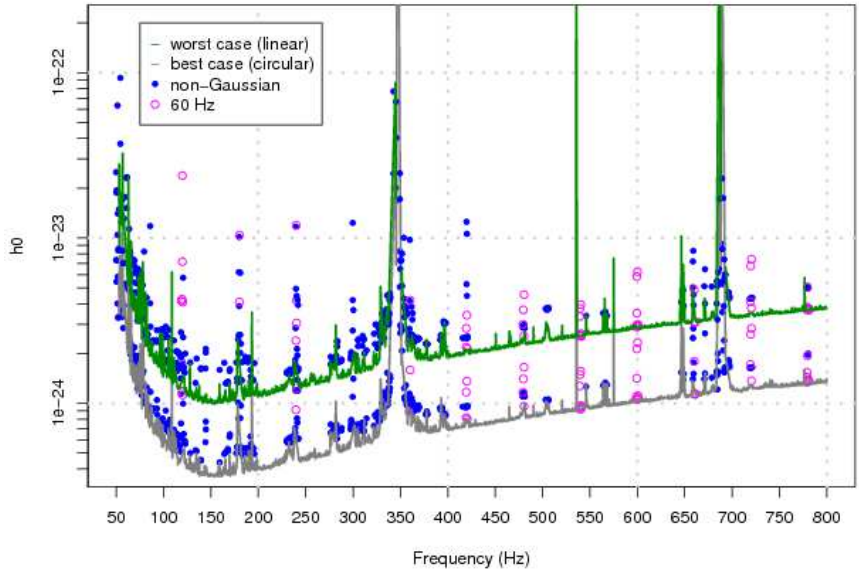}
  \caption{
All-sky upper limits on the GW amplitude of unknown sources.
The upper (green) curve shows worst-case upper limits (most unfavorable orientation of
a linearly polarized source) in analyzed 0.25~Hz bands.
The lower (gray) curve shows upper limits assuming a circularly polarized source.
Solid points and circles denote bands with severe 
instrumental contaminations and hence unreliable limits.}
    \label{fig:S5targeted}
 \end{center}
 \end{figure}

Because of computational costs of searching long observations times without
{\it a priori} knowledge of gravitational wave frequency evolution, one must
make tradeoffs in directed searches for particular objects or points in the
sky. A published broadband search (see \cite{bib:casa}) for the X-ray-emitting compact central object in the
Cassiopeia supernova remnant provides one example, based on analysis of a subset of
LIGO S5 data (time span of $\sim$12 days), for which the lowest strain upper limit
was $7\times10^{-25}$ at $\sim$150 Hz.
Because Cas A is only about 300 years old, this search incorporated
a search over spin frequency and over its 1st and 2nd time derivatives. The resulting large
parameter space volume for even a single point on the sky led to degraded strain sensitivity,
compared to that achieved in the targeted searches described above.


All-sky searches for unknown neutron stars must cope with a still larger parameter space
volume (as quantified by number of distinct templates searched for a fixed maximum SNR mismatch).
Figure~\ref{fig:S5targeted} shows all-sky strain upper limits~\cite{bib:cwpowerfluxs5} 
on spinning isolated neutron
stars,
based on analysis of the full S5 data set, using semicoherent sums of
Doppler-demodulated Fourier transform powers from tens of thousands of half-hour intervals 
(``PowerFlux'' algorithm).
A complementary and wider-band search of S5 data, based on 
Fourier transforms of longer coherence time (up to 25 hours per interferometer) 
and using the Einstein@Home distributed-computing project,
led to comparable sensitivity in a search recently submitted for 
publication (see \cite{bib:cweinsteinathomes5}). 


Comparison of the targeted and all-sky strain upper limits shown in figure~\ref{fig:S5targeted}
confirms the expected (substantial) degradation of sensitivity for searches that must search large parameter space
volumes and hence must set high SNR thresholds, to cope with otherwise increased statistical outlier counts.
For this reason, the ``photon astronomer'' community is encouraged not only to search for new and exotic 
objects that could serve as potential gravitational wave candidates, \eg, nearby pulsars with high
rotational energy losses, but also to determine spin rotations for known objects, such as Cas A
and Scorpius X-1. 

Installation of Advanced LIGO and Advanced Virgo has begun, with early operation
expected circa 2015. When these detectors reach design sensitivity near the end of the decade,
strain amplitude sensitivities and hence ranges within the galaxy will improve by an order of magnitude.
Electromagnetic measurements could well make the difference between discovering and missing a star
with a detectable gravitational wave signal.

\acknowledgements
The authors gratefully acknowledge the support of the United States
National Science Foundation for the construction and operation of the
LIGO Laboratory, the Science and Technology Facilities Council of the
United Kingdom, the Max-Planck-Society, and the State of
Niedersachsen/Germany for support of the construction and operation of
the GEO600 detector, and the Italian Istituto Nazionale di Fisica
Nucleare and the French Centre National de la Recherche Scientifique
for the construction and operation of the Virgo detector. The authors
also gratefully acknowledge the support of the research by these
agencies and by the Australian Research Council, 
the International Science Linkages program of the Commonwealth of Australia,
the Council of Scientific and Industrial Research of India, 
the Istituto Nazionale di Fisica Nucleare of Italy, 
the Spanish Ministerio de Econom\'ia y Competitividad,
the Conselleria d'Economia Hisenda i Innovaci\'o of the
Govern de les Illes Balears, the Foundation for Fundamental Research
on Matter supported by the Netherlands Organisation for Scientific Research, 
the Polish Ministry of Science and Higher Education, the FOCUS
Programme of Foundation for Polish Science,
the Royal Society, the Scottish Funding Council, the
Scottish Universities Physics Alliance, The National Aeronautics and
Space Administration, 
the National Research Foundation of Korea,
Industry Canada and the Province of Ontario through the Ministry of Economic Development and Innovation, 
the National Science and Engineering Research Council Canada,
the Carnegie Trust, the Leverhulme Trust, the
David and Lucile Packard Foundation, the Research Corporation, and
the Alfred P. Sloan Foundation.

\end{document}